\documentclass{article}

\usepackage{arxiv}

\usepackage[utf8]{inputenc}
\usepackage[super,comma,round]{natbib}
\usepackage[figuresleft]{rotating}
\usepackage{txfonts}%
\usepackage{multirow}
\usepackage{array}
\usepackage{xcolor}
\usepackage[hidelinks]{hyperref}

\newcolumntype{L}[1]{>{\raggedright\let\newline\\\arraybackslash\hspace{0pt}}m{#1}}
\newcolumntype{C}[1]{>{\centering\let\newline\\\arraybackslash\hspace{0pt}}m{#1}}
\newcolumntype{R}[1]{>{\raggedleft\let\newline\\\arraybackslash\hspace{0pt}}m{#1}}

\title{Integrated Dataset of Brazilian Flights}

\author{
 Claudio~Teixeira \\
 CEFET/RJ\\
 \texttt{claudio.teixeira@eic.cefet-rj.br} \\
 \And
 Lucas~Giusti \\
 CEFET/RJ\\
 \texttt{lucas.giusti@eic.cefet-rj.br} \\
  \AND
 Jorge~Soares\\
 CEFET/RJ\\
 \texttt{jorge.soares@cefet-rj.br} \\
  \And
Joel~dos~Santos\\
 CEFET/RJ\\
 \texttt{joel.santos@cefet-rj.br} \\
  \AND
Glauco~Amorim\\
 CEFET/RJ\\
 \texttt{glauco.amorim@cefet-rj.br} \\
  \And
Eduardo~Ogasawara\\
 CEFET/RJ\\
 \texttt{eogasawara@ieee.org} \\
}

\begin{document}
\maketitle

\begin{abstract}
The Brazilian commercial aviation system achieved the first position among Latin American countries and the fifteenth place worldwide on the Revenue Passenger-Kilometer (RPK) ranking. The availability of data regarding flight, including flight information and meteorological conditions, enables studies about the Brazilian flight system, such as flight delays and timetabling. Therefore, this paper contributes to such studies by offering an integrated dataset containing data on departure and arrival for flights departing and arriving on Brazilian airports comprising the period from $2000$ to $2019$. This paper presents a dataset composed of $15,505,922$ records of flight data, each containing $45$ attributes. The attributes include data regarding the airline, flight, airports, meteorological conditions, scheduled and elapsed times for departure and arrival.

\end{abstract}

\section{Introduction}
\label{sec:intro}

The Brazilian commercial aviation system contains more than one hundred airports. It transported 95.9 million revenue passengers during $2014$. It achieved the first position among Latin American countries and the fifteenth place worldwide on the Revenue Passenger-Kilometer (RPK) ranking \citep{icao_annual_2015}. The commercial aviation network in Brazil is organized towards regional hubs in contrast to airline hubs. The main reason is the Brazilian territorial extension and that few Brazilian states have more than one major airport. One exception to this rule is Campinas (in the state of São Paulo), where airline company \textit{Azul} holds $77\%$ of its commercial flights. Besides, due to market deregulation instituted in $2005$, the Brazilian commercial aviation system experienced significant changes in its players, leading to market share changes and flight availability.

The National Civil Aviation Agency (ANAC) is responsible for regulating and supervising the Brazilian civil aviation activities. Since $2000$, ANAC keeps track of departure and arrival data for Brazilian flights in its Active Regular Flight (VRA) dataset \citep{anac_agencia_2015}. The data available in VRA are registered by the airlines and consolidated by ANAC. It contains data about each flight stage, \emph{i.e.}, the aircraft's necessary steps from its takeoff to the next landing. These steps are established regardless of where the object of transport has been loaded or unloaded. For each flight step, VRA provides data such as airline, flight number, type (such as international, domestic, and cargo), class (such as regular, extra, charter, and instruction), airports, and scheduled and elapsed times for departure and arrival. ANAC monthly provides VRA data on its webpage.

VRA enables studying the Brazilian commercial aviation system. Examples of studies are flight delay patterns \citep{sternberg_experimental_2016} and their prediction \citep{moreira_evaluating_2018,scarpel_data_2018}. Although meteorological conditions play an essential role in analyzing flight information, such data is not present in VRA. Thus, this paper presents a dataset that integrates Brazilian flight data. It fuses all monthly data available in VRA. It enriches it with meteorological data from the ASOS (Automated Surface Observing Systems) dataset \citep{asos_automated_2000} provided by the IOWA University in the USA. ASOS contains weather sensor data from airports around the world. During the entire data integration, data cleaning and data preprocessing techniques were also applied to improve its quality.

\section{Data Acquisition}
\label{sec:acquisition}

According to the flight regulation of ANAC, commercial airline companies must register flight metadata indicating changes in flight time, either delay, anticipation, or canceling. They have to log the time a flight happened and a justification for the alteration. Table~\ref{tab:vra} indicates the flight metadata together with their semantics.

\begin{table}[!ht]
	\caption{Flight metadata registered by airline companies available in VRA}
	\centering
	\begin{tabular}{l p{8cm}}
		\textbf{Attribute} & \textbf{Description} \\
		\hline
		Airline & ICAO code representing the airline company \\
		Flight & Flight number \\
		Authorization code & Identifies the authorization type for each flight step \\
		Flight type & Identifies the type of operation performed \\
		Origin & ICAO code of origin airport \\
		Destination & ICAO code of destination airport \\
		Expected Departure & Date and time of scheduled departure \\
		Real departure & Date and time of departure performed informed by the airline \\
		Estimated Arrival & Date and time of estimated arrival \\
		Real Arrival & Date and time of arrival, informed by the airline \\
		Flight status & Informs if the flight was performed or canceled \\
		Justification Code & Identifies the delay, cancellation, and other changes concerning the planned flight
	\end{tabular}
	\label{tab:vra}
\end{table}

According to the regulation of ANAC, the metadata indicated in Table 2 must be registered in a paper form, either typed or handwritten. ANAC then consolidate the data sent by the airline companies into the VRA dataset. VRA is published monthly, comprising all flight steps expected to depart in a given month.

The primary goal of ANAC is to use the recorded metadata to compute the punctuality rate of airlines. Thus, sector regulation obliges airline companies to provide the data presented in Table~\ref{tab:vra}. Therefore, it comprises all flight steps that took place in a given period. However, around 20\% of the records may be considered inconsistent due to errors while filling the report form. As will be presented in Section~\ref{sub:preprocessing}, the causes of errors include arrival time before departure or flight duration inconsistent with the regulation of ANAC.

\begin{figure}[!ht]
	\centering
	\includegraphics[width=.65\textwidth]{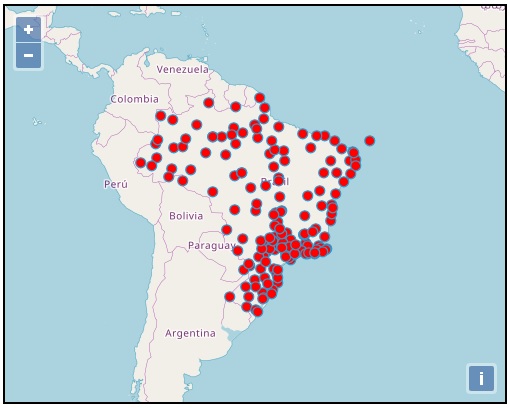}
	\caption{Brazilian airports included in the ASOS dataset \citep{asos_automated_2000}}
	\label{fig:asos_airports}
\end{figure}

Meteorological conditions play an important role in aviation operations. The Automated Surface Observing Systems (ASOS) is a program that involves several American government agencies. It was created to become an official network of meteorological information to support primarily aviation entities. It includes meteorological, climatological, and hydrological components. ASOS data come from weather sensors in locations all over the planet. In Brazil, ASOS covers all $154$ airports available in VRA, as seen in Figure~\ref{fig:asos_airports}.

The Department of Agronomy at Iowa State University, in the United States, compiles daily information from the US ASOS system. It creates an hourly report of meteorological observations in all of its sites. Table~\ref{tab:asos} indicates the meteorological data together with their semantics.

\begin{table}[!ht]
	\caption{ASOS meteorological data}
	\centering
	\begin{tabular}{l p{8cm}}
		\textbf{Attribute} & \textbf{Description} \\
		\hline
		Sky condition & Cloud height and amount (clear, scattered, broken, overcast) up to 12,000 feet \\
		Visibility & To at least ten statute miles \\
		Weather & Type and intensity for rain, snow, and freezing rain. \\
		Obstructions to vision & fog, haze\\
		Pressure & Sea-level pressure, altimeter setting\\
		Temperature & Ambient and dew point temperature\\
		Wind & Direction, speed, and character (gusts, squalls)\\
		Precipitation accumulation\\
	\end{tabular}
	\label{tab:asos}
\end{table}

\section{Integrated Dataset}
\label{sec:data}

The integrated \emph{Brazilian Flight Dataset} (BFD) presented in this paper includes both the flight data present in VRA and meteorological information present in ASOS. It is intended to enable studies regarding the Brazilian commercial aviation system. BFD is composed of $15,505,922$ records of flight data, each containing $45$ attributes. The dataset, together with its integration process description and R scripts, is available on IEEE DataPort\footnote{Dataset is available at \url{http://dx.doi.org/10.21227/k10b-qn21}. Additional information can be found at \citep{teixeira_integrated_2020}.}.

\begin{figure}[!ht]
	\centering
	\includegraphics[width=.8\textwidth]{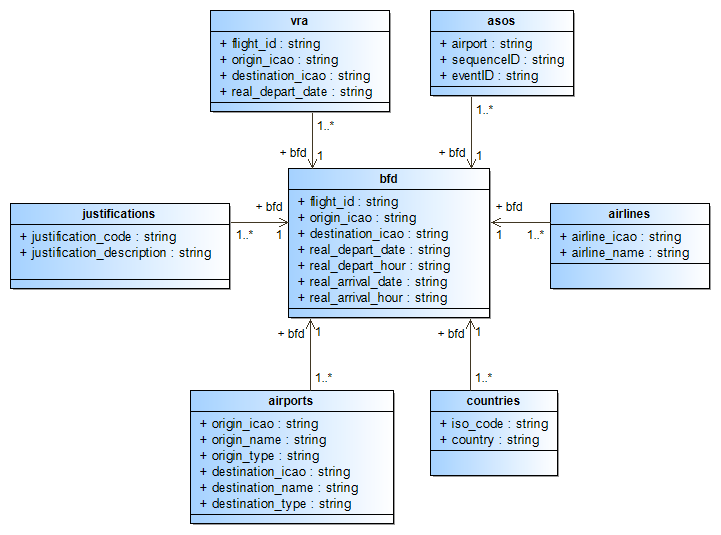}
	\caption{The data model for the BFD}
	\label{fig:bfdmodel}
\end{figure}

Figure~\ref{fig:bfdmodel} presents the data model of BFD. It is detailed in the following sections. As can be seen, BFD aggregates data from VRA and ASOS for flight information and meteorological information, respectively. It also includes data currently unavailable in VRA, such as describing the justification codes of ANAC, airline and airport names, and ISO codes for country names.

BFD focus on flight data regarding flights that departed or arrived in Brazil. When both origin and destination airports are located in Brazil, those flights are considered domestic flights. Conversely, when either the origin or the destination airport is located outside of Brazil, it is considered international. The data integration process for creating BFD was organized into three main activities: (i) data preprocessing, (ii) data enrichment, and (iii) data fusion. Those activities resemble the traditional Extraction, Transformation, and Load (ETL) process \citep{vassiliadis_survey_2009}.

\subsection{Data Preprocessing}
\label{sub:preprocessing}

The preprocessing stage was performed in three parts. First, VRA attribute names were translated from Brazilian Portuguese to English. It was unnecessary to translate the acronyms used in each variable since they were already following the International Civil Aviation Organization (ICAO) standards. It was necessary to convert temperature and dew point data to the International System of Units regarding the ASOS data. Data from ASOS was filtered to consider the $154$ airports available in VRA.

The second part consisted of data cleaning for both VRA and ASOS datasets.
Given that flight information is usually recorded by hand, VRA data was cleaned to remove inconsistent data. During cleaning, records with missing variables were removed. Also, records with departure time (either elapsed or expected) greater or equal to arrival time were removed. They corresponded to approximately $0.02\%$ of the records. Approximately $3.77\%$ of VRA records were removed for being out of BFD scope, \emph{i.e.}, with origin and destination out of Brazil. Finally, the regulation of ANAC prohibits delays higher than $24$ hours. Thus, during cleaning records with departure or arrival delays exceeding this norm were removed. The complete data cleaning removed $21.07\%$ of VRA records.

The third part of the preprocessing stage consisted of removing outliers. For each pair of airports $\langle o,d\rangle$ in VRA, it was considered both the expected and elapsed duration of a flight from origin $o$ and destination $d$. Flights whose duration (either elapsed or expected) were not in the interval $[Q_1 - 3 \cdot IQR, Q_3 + 3 \cdot IQR]$ were considered as outliers. They corresponded to 2.76\% of VRA records. The preprocessing step resulted in $15,505,922$ flight records from VRA to be used in the fusion stage.

\subsection{Data Enrichment}
\label{sub:enrichment}

After preprocessing, the dataset is enriched as follows. The dataset schema is changed by separating departure and arrival data attributes (see Table~\ref{tab:vra} into an hour and date attributes. Besides, it included attributes related to flight duration, departure and arrival delays.

Additionally, two discrete attributes were included for the time of the day for departures and arrivals. It divides the time attribute into seven ranges, as presented in Table~\ref{tab:periodos}.

\begin{table}[!ht]
	\caption{Time attribute discretization}
	\centering
	\begin{tabular}{l c c}
		Period & Start Time & End Time \\
		\hline
		Night & 23:00 & 04:00 \\
		Early Morning & 05:00 & 08:00 \\
		Mid Morning & 09:00 & 10:00 \\
		Late Morning & 11:00 & 12:00 \\
		Afternoon & 13:00 & 16:00 \\
		Early Evening & 17:00 & 19:00 \\
		Late Evening & 20:00 & 22:00 \\
	\end{tabular}
	\label{tab:periodos}
\end{table}

Two discrete attributes are included in ASOS while enriching the dataset. The use the wind velocity in knots to include the wind intensity using a Beaufort Scale. The second uses the wind direction in degrees to include the wind direction using Wind Rose with 16 cardinal directions (N, NNE, NE, ENE, E, ESE, SE, SSE, S, SSW, SW, WSW, W, WNW, NW and NNW)\footnote{Wind Rose Data - US Department of Agriculture - Natural Resources Conservation Service (NRCS) available at \url{https://www.wcc.nrcs.usda.gov/climate/windrose.html}}.

\subsection{Data Fusion}
\label{sub:fusion}

Data fusion was applied over VRA data from 2000 to 2019, except for June, July 2014, and March 2018, when ANAC did not collect the data. It is worth mentioning that ASOS provides hourly meteorological data.  

During the fusion process for the meteorological and flight data, it was necessary to group all flight data in a given hour. The grouping was performed for each elapsed departure and arrival of the flight to determine its meteorological information. 

Furthermore, the fusion stage resolved airport and airline names from VRA data. It also included an ISO code for country names whenever the flight departs or arrives at a non-Brazilian airport. Finally, the justification codes for flight delay were also expanded to their descriptions.

\section{Dataset Usage}

BFD allows for studies regarding the Brazilian commercial aviation system. In this section, we present previous and ongoing work conducted on top of BFD together with an exploratory analysis of BFD data. To present the importance of using the database, we conduct an exploratory analysis and mention studies that used the data in their research.

\begin{figure}[!ht]
	\centering
	\includegraphics[width=.8\textwidth]{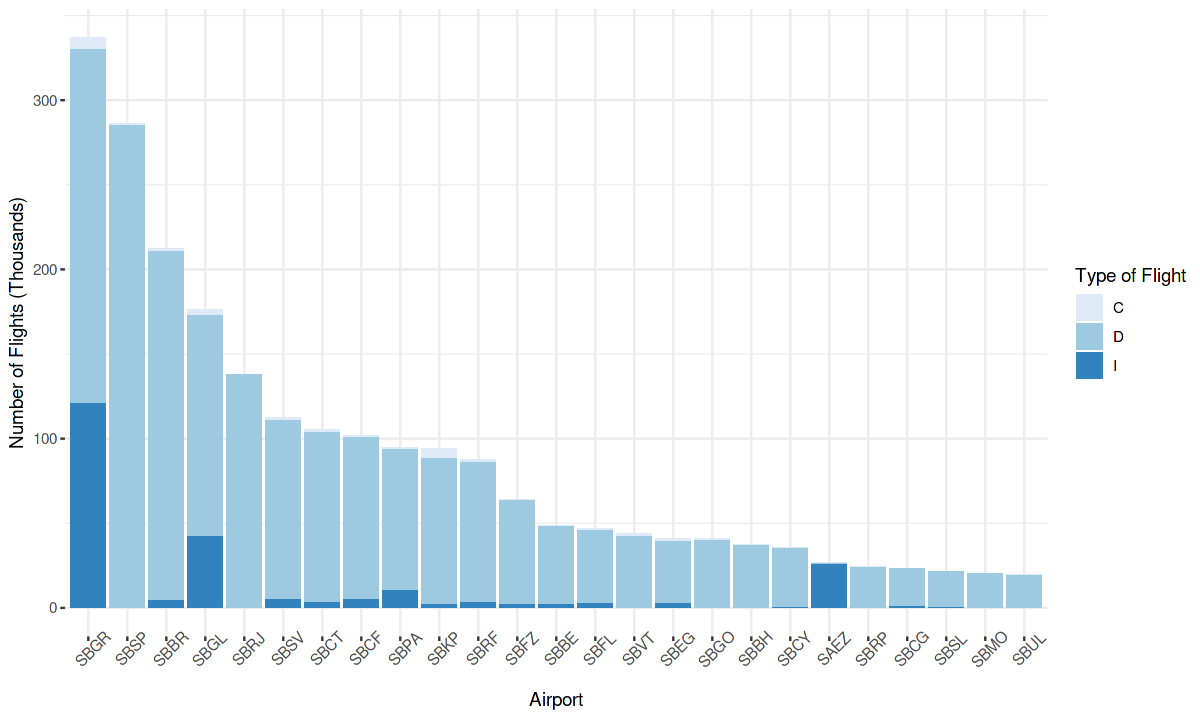}
	\caption{Number of flights per airport, for the top-25 most active airports}
	\label{fig:voos_aeroporto}
\end{figure}

As discussed before, the Brazilian flight system is oriented towards regional hubs instead of company hubs. Figure~\ref{fig:voos_aeroporto} presents the number of flights per airport, considering just the 25 biggest airports on flights. It also divides flights into domestic (D), international (I), and cargo (C) flights. 

As can be seen in Figure~\ref{fig:voos_aeroporto}, in the top five busiest airports, the first two are in São Paulo (SBSP and SBGR), the third in Brasília (SBBR), and the last two in Rio de Janeiro (SBGL and SBRJ). Rio and São Paulo are the two higher Gross Domestic Products (GDPs) in Brazil. They are two major gateways for flights coming and exiting Brazil. Approximately one-third of the flight in Guarulhos Airport (SBGR) and Galeão Airport (SBGL) are international flights.

Brasilia is the capital of the country and is located in the middle of Brazil. It acts as a hub for flights from and to cities in the north and northeast regions. It can be seen, however, that it has few international flights.

Brazil and Argentina have strong touristic relations. Thus we can see the Buenos Aires international airport (SAEZ) in the top-25 busiest airports. Since BFD has only flights from and to Brazil, SAEZ has only international and cargo flights.

\begin{figure}[!ht]
	\centering
	\includegraphics[width=.8\textwidth]{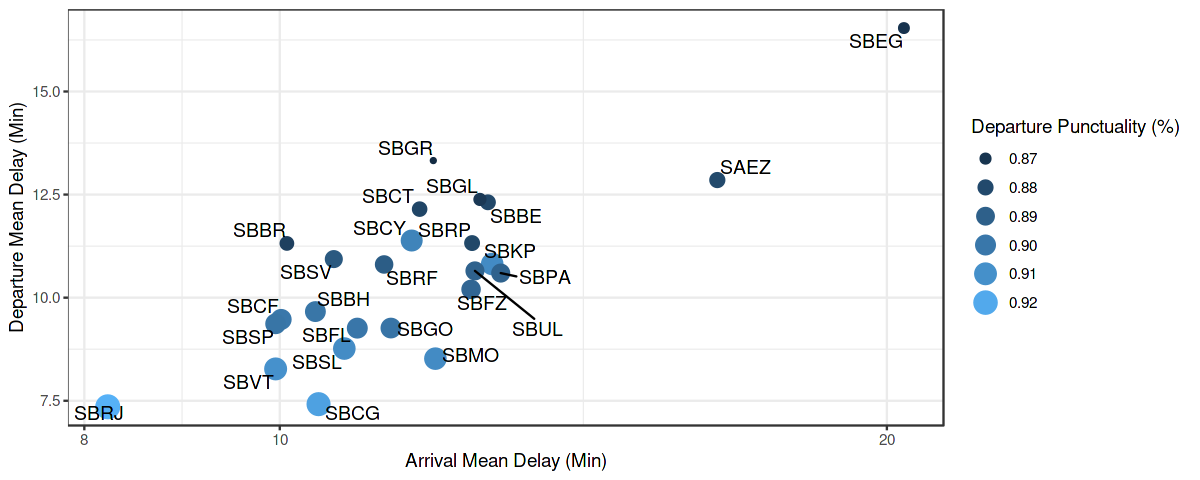}
	\caption{Mean takeoff delay and punctuality rate per mean arrival delay for the top-25 busiest airports}
	\label{fig:atraso_aeroporto}
\end{figure}

Figure~\ref{fig:atraso_aeroporto} presents the takeoff and arrival delay per airport for the top-25 busiest airports. It indicates whether an airport has recover capabilities for arrival delays. The radius of the airport also indicates the level of punctuality. The higher the radius, the airports are more punctual. 

\begin{figure}[!ht]
	\centering
	\includegraphics[width=.9\textwidth]{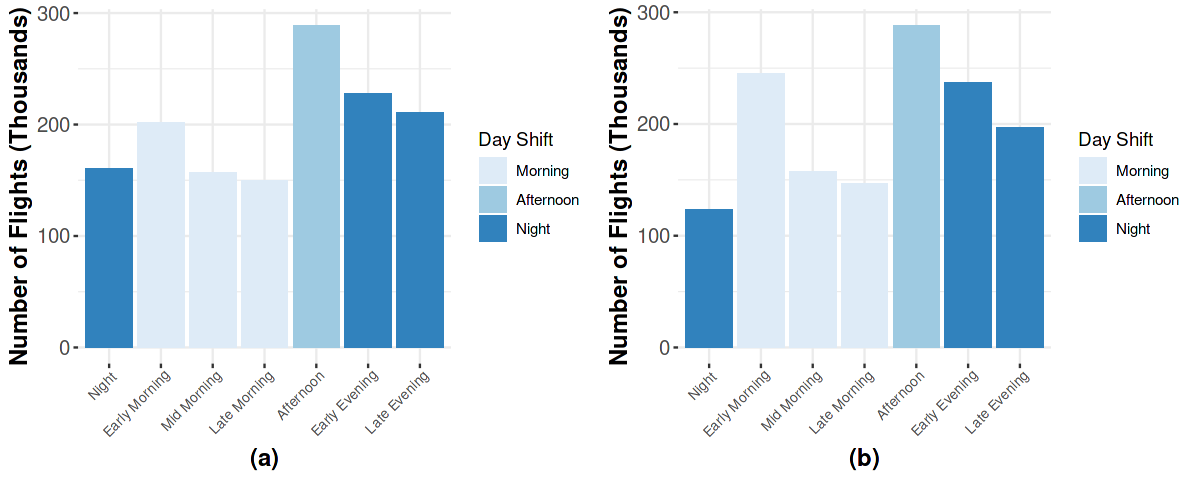}
	\caption{Number of flights per period of the day: (a) departure; (b) arrival}
	\label{fig:voos_horario}
\end{figure}

Figure~\ref{fig:voos_horario} presents the distribution of flights according to the period of the day. As shown, most of the flight departures (Figure~\ref{fig:voos_horario}.a) occur in the afternoon and early evening. Most arrivals (Figure~\ref{fig:voos_horario}.b) occur in the afternoon and early morning. During the mid and late morning, the number of flights decreases significantly for both departure and arrival. 

\begin{figure}[!ht]
	\centering
	\includegraphics[width=.95\textwidth]{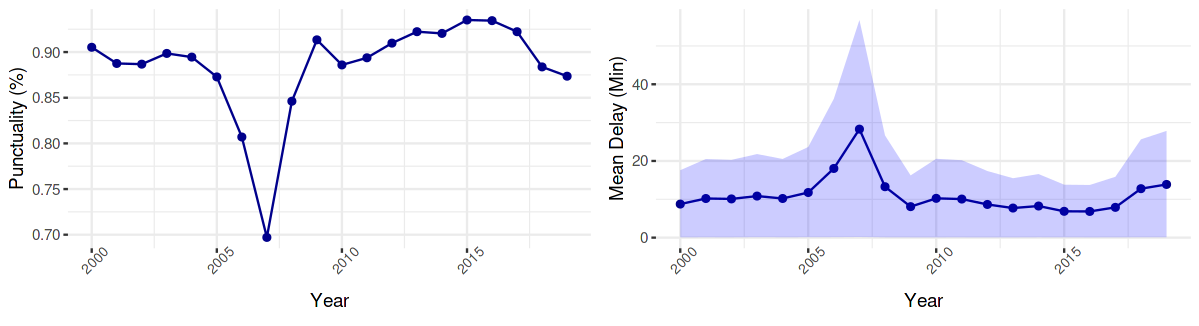}
	\caption{Punctuality rate and mean delay per year. The charts present the mean delay together with its confidence interval of $95\%$}
	\label{fig:pontualidade_tempo}
\end{figure}

According to ANAC regulation, a flight is considered to be delayed when its departure or arrival time surpasses, respectively, the expected departure or arrival by more than $30$ minutes. Figure~\ref{fig:pontualidade_tempo} presents the punctuality rate considering all the Brazilian flight systems per year, from $2000$ to $2019$. It is possible to observe that the Brazilian flight crises that occurred in $2007$ interfered with both punctuality rates and mean delay \citep{times_brazil_2007}.

\begin{figure}[!ht]
	\centering
	\includegraphics[width=.95\textwidth]{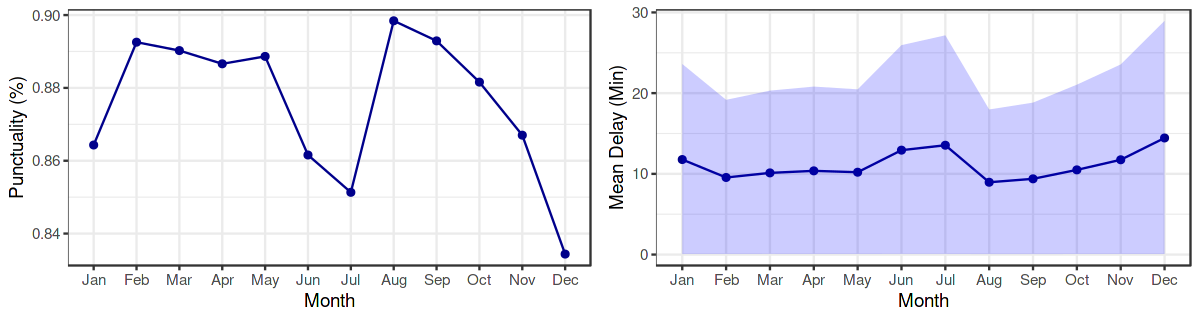}
	\caption{Punctuality rate and mean delay per month of the year}
	\label{fig:pontualidade_tempo_mes}
\end{figure}

Figure~\ref{fig:pontualidade_tempo_mes} analysis of the Brazilian systems monthly. Historically, months of school break (December, January, and July) have the lowest punctuality rates and the highest mean delay. August is the month with the highest level of punctuality and lowest mean delay. 

\begin{figure}[!ht]
	\centering
	\includegraphics[width=.8\textwidth]{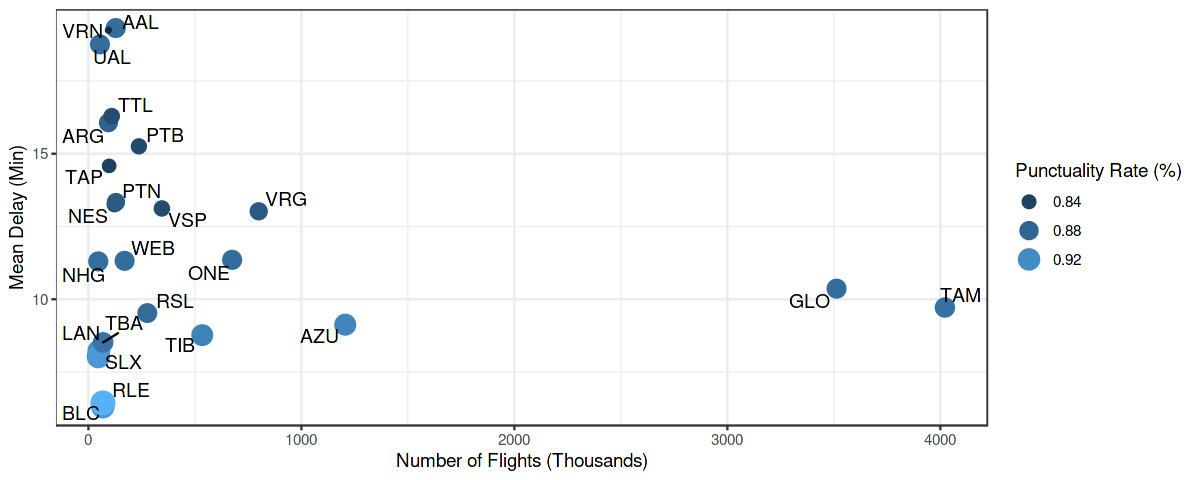}
	\caption{Mean delay and Punctuality rate per number of flights for the top-25 airline companies}
	\label{fig:empresas}
\end{figure}

Finally, Figure~\ref{fig:empresas} presents the punctuality rate (circle size) and the average delay in minutes per number of flights for the top-25 companies. According to Figure~\ref{fig:empresas}, two airlines present the most significant number of flights, TAM and Gol (GLO). It is also possible to observe that airlines with lower punctuality rates tend to have a higher mean delay.

Given the various inconveniences for airlines, airports, and passengers caused by flight delays,  it is fundamental to mitigate their occurrence and optimize an air transport system's decision-making process. Mainly, airlines, airports, and users may be more interested in when delays are likely to occur than the accurate prediction of the absence of delays. In that context, \citet{moreira_evaluating_2018} use BFD to analyze Flight delays in the period between $2009$ and $2015$. The authors present a classification model capable of predicting delays, getting about $60\%$ of hits.

Flight delays fall into two main categories: root delay and delay propagation. Root delays are related to events that are intrinsic to a particular flight. In delay propagation, it is presumed that a delay has already occurred at some point in the network, \emph{i.e.}, new delays occur due to previous delays. The understanding of delay propagation patterns among airports is essential for decision-making processes.

That study may devise patterns in flight delays and the way the system recover from it. Focusing on unveiling those patterns, \citet{sternberg_analysis_2016} apply data indexing techniques combined with BFD data association rules. The authors observed that the Brazilian flight system has difficulties recovering from previous delay when operating under adverse meteorological conditions, when delays occurrences may increase up to $216\%$.

\section{Conclusion}

This work aimed to create a reliable and enriched database on national and international flights that arrived and departed from Brazilian airports. With the data offered by this database, it is possible to carry out several studies to aid the decision-making process. For example, it is possible to answer the following questions: (i) ``Which airport suffers the most delays?''; (ii) ``What month of the year is an airport most likely to be delayed?''; or (iii) ``What part of the day is a particular airport most likely to experience a delay in departure?''

The answers to these questions can help companies and governments review their protocols and optimize their services. Additionally, we intend to update this dataset yearly, conducting the entire data integration.  

\section*{Acknowledgments}
The authors thank CNPq, CAPES (finance code 001), FAPERJ, and CEFET/RJ for partially funding this research.

\section*{Conflict of interest}
On behalf of all authors, the corresponding author states that there is no conflict of interest.

\section*{Author's contributions}
All authors contributed equally to the study. EO conceptualized the study design. CT acquired the data. LT and JS conducted data analysis and interpretation. Furthermore JAS and GA revised it critically for intellectual content. All authors have approval of the final version.

\bibliographystyle{abbrvnat}
\bibliography{references}

\begin{thebibliography}{10}
\providecommand{\natexlab}[1]{#1}
\providecommand{\url}[1]{\texttt{#1}}
\expandafter\ifx\csname urlstyle\endcsname\relax
  \providecommand{\doi}[1]{doi: #1}\else
  \providecommand{\doi}{doi: \begingroup \urlstyle{rm}\Url}\fi

\bibitem[{ANAC}(2015)]{anac_agencia_2015}
{ANAC}.
\newblock Agência {Nacional} de {Aviação} {Civil}.
\newblock Technical report, \url{https://www.gov.br/anac/pt-br}, 2015.

\bibitem[{ASOS}(2000)]{asos_automated_2000}
{ASOS}.
\newblock Automated {Surface} {Observing} {System}.
\newblock Technical report, \url{https://mesonet.agron.iastate.edu/ASOS/},
  2000.

\bibitem[{ICAO}(2015)]{icao_annual_2015}
{ICAO}.
\newblock Annual {Report} of the {Council} 2014.
\newblock Technical report,
  \url{http://www.icao.int/annual-report-2014/Pages/default.aspx}, 2015.

\bibitem[Moreira et~al.(2018)Moreira, Dantas, Oliveira, Soares, and
  Ogasawara]{moreira_evaluating_2018}
L.~Moreira, C.~Dantas, L.~Oliveira, J.~Soares, and E.~Ogasawara.
\newblock On {Evaluating} {Data} {Preprocessing} {Methods} for {Machine}
  {Learning} {Models} for {Flight} {Delays}.
\newblock In \emph{Proceedings of the {International} {Joint} {Conference} on
  {Neural} {Networks}}, volume 2018-July, 2018.

\bibitem[Scarpel and Pelicioni(2018)]{scarpel_data_2018}
R.~A. Scarpel and L.~Pelicioni.
\newblock A data analytics approach for anticipating congested days at the
  {São} {Paulo} {International} {Airport}.
\newblock \emph{Journal of Air Transport Management}, 72:\penalty0 1--10, 2018.

\bibitem[Sternberg et~al.(2016{\natexlab{a}})Sternberg, Carvalho, Murta,
  Soares, and Ogasawara]{sternberg_analysis_2016}
A.~Sternberg, D.~Carvalho, L.~Murta, J.~Soares, and E.~Ogasawara.
\newblock An analysis of {Brazilian} flight delays based on frequent patterns.
\newblock \emph{Transportation Research Part E: Logistics and Transportation
  Review}, 95:\penalty0 282--298, 2016{\natexlab{a}}.

\bibitem[Sternberg et~al.(2016{\natexlab{b}})Sternberg, Carvalho, Murta,
  Soares, and Ogasawara]{sternberg_experimental_2016}
A.~Sternberg, D.~Carvalho, L.~Murta, J.~Soares, and E.~Ogasawara.
\newblock Experimental {Evaluation}.
\newblock Technical report,
  \url{https://eic.cefet-rj.br/~dal/an-analysis-of-brazilian-flight-delays-based-on-frequent-patterns/},
  2016{\natexlab{b}}.

\bibitem[Teixeira et~al.(2020)Teixeira, Teixeira, dos Santos, Amorim, Soares,
  and Ogasawara]{teixeira_integrated_2020}
C.~Teixeira, L.~Teixeira, J.~dos Santos, G.~Amorim, J.~Soares, and
  E.~Ogasawara.
\newblock Integrated {Brazilian} {Flight} {Datasets} {Description}.
\newblock Technical report,
  \url{https://eic.cefet-rj.br/~dal/brazilian-flight-dataset-description},
  2020.

\bibitem[Times(2007)]{times_brazil_2007}
N.~Y. Times.
\newblock Brazil {Demands} {Solution} to {Aviation} {Crisis}.
\newblock Technical report,
  \url{https://www.nytimes.com/2007/07/19/world/americas/19brazil.html}, 2007.

\bibitem[Vassiliadis(2009)]{vassiliadis_survey_2009}
P.~Vassiliadis.
\newblock A survey of extract-transform-load technology.
\newblock \emph{International Journal of Data Warehousing and Mining},
  5\penalty0 (3):\penalty0 1--27, 2009.

\end{thebibliography}

\end{document}